\documentclass[twocolumn,showpacs,preprintnumbers,amssymb]{revtex4-1}
\usepackage{graphicx}
\usepackage{dcolumn}
\usepackage{color}
\usepackage{bm}

\def \be{\begin{equation}}
\def \ee{\end{equation}}
\def \ben{\begin{eqnarray}}
\def \een{\end{eqnarray}}

\begin{document}

\title{The Hawking Temperature in the context of Dark Energy for
  Reissner-Nordstrom and Kerr background}

\author{Goutam Manna }
\altaffiliation{goutammanna.pkc@gmail.com}
\affiliation{Department of Physics, Prabhat Kumar College, Contai,Purba Medinipur-721401,India}

\author{Debashis Gangopadhyay}
\altaffiliation{debashis@rkmvu.ac.in}
\affiliation{Department of Physics, Ramakrishna Mission Vivekananda University, P.O.-Belur Math, Howrah-711202, West Bengal, India.}

\begin{abstract}
For emergent gravity metrics, presence of dark energy modifies the Hawking temperature. We show that for 
the spherically symmetric Reissner-Nordstrom (RN) background metric, 
the emergent metric can be mapped into a Robinson-Trautman blackhole. Allowed values of 
the dark energy density follow from rather general conditions. For some allowed value of the dark energy 
density this blackhole 
can have zero Hawking temperature i.e. the blackhole does not radiate. 
For a  Kerr background along $\theta=0$ , 
the emergent blackhole metric satisfies Einstein's equations for large $r$ and always radiates. Our analysis is done in the 
context of emergent gravity metrics having $k-$essence scalar fields $\phi$ with a Born-Infeld type lagrangian. 
In both cases the 
scalar field $\phi(r,t)=\phi_{1}(r)+\phi_{2}(t)$ also satisfies the emergent gravity equations of motion 
for $r\rightarrow\infty$ and $\theta=0$. 

\keywords{dark energy, k-essence ,Reissner-Nordstrom and Kerr blackholes}

\pacs{98.80.-k ;95.36.+x}

\end{abstract}
\maketitle

\section{Introduction}
In \cite{dg} it has been shown that the Hawking temperature \cite{haw} is modified in the
presence of dark energy. In \cite{dg} this was shown for an emergent gravity metric 
$\tilde G_{\mu\nu}$
having $k-$ essence scalar fields $\phi$ with a Born-Infeld type lagrangian and with the gravitational 
metric as Schwarzschild. The Lagrangian for $k-$essence scalar fields contain non-canonical kinetic terms. 
The general form for such  lagrangians is proportional to $F(X)$ with 
$X={1\over 2}g^{\mu\nu}\nabla_{\mu}\phi\nabla_{\nu}\phi$.
Relevant literature for such fields in  cosmology, inflation, dark matter, dark energy 
and strings can be found in \cite{scherrer}. 

The motivation of this work is to calculate the Hawking temperature   
for an emergent gravity metric in the presence of dark energy  
and which is also a blackhole metric.We consider two cases, i.e.,
when the gravitational metric is a (a) Reissner-Nordstrom blackhole metric 
and (b) Kerr blackhole metric. 
As explained in detail in section 2, $\tilde G_{\mu\nu}$ contains the 
dark energy field $\phi$ and this should satisfy the emergent gravity equations of motion. 
Again, for  $\tilde G_{\mu\nu}$ to be a blackhole metric, it has to satisfy the Einstein field equations. 
In Ref. \cite{dg} this was shown by mapping the emergent gravity metric (having Schwarzschild background) 
into a Barriola-Vilenkin blackhole metric which again satisfies the Einstein equations.
Here we find that for the RN background  case the emergent gravity metric can be exactly mapped onto a 
Robinson-Trautman blackhole so that the Einstein equations are automatically satisfied. However,the 
$k$-essence matter fields satisfy the emergent gravity equations of motion only for $\theta=0$. For the 
Kerr case, the emergent metric  satisfies Einstein equations for  large $r$ while the  dark energy field 
$\phi$ satisfies the emergent gravity equations of motion again only for $\theta=0$.  

In this context we clarify that the Hawking temperature is spherically symmetric from very general conditions 
and taking $\theta=0$ does not therefore affect this property of the Hawking temperature.  
We will elaborate on this in a more quantitative way in sections 3 and 5.
The formalism for  emergent gravity used is as described in \cite{babi}.

\section{Emergent Gravity}
The emergent gravity minimal action for background metric $g_{\mu\nu}$ is 
\ben
S_{k}[\phi,g_{\mu\nu}]= \int d^{4}x {\sqrt -g} L(X,\phi)
\label{eq:1}
\een
The energy momentum tensor is
\ben
T_{\mu\nu}\equiv {2\over \sqrt {-g}}{\delta S_{k}\over \delta g^{\mu\nu}}= L_{X}\nabla_{\mu}\phi\nabla_{\nu}\phi - g_{\mu\nu}L
\label{eq:2}
\een
$L_{\mathrm X}= {dL\over dX},~~ L_{\mathrm XX}= {d^{2}L\over dX^{2}},
~~L_{\mathrm\phi}={dL\over d\phi}$ and  
$\nabla_{\mu}$ is the covariant derivative with respect to the metric $g_{\mu\nu}$.
The equation of motion is
\ben
-{1\over \sqrt {-g}}{\delta S_{k}\over \delta \phi}= \tilde G^{\mu\nu}\nabla_{\mu}\nabla_{\nu}\phi +2XL_{X\phi}-L_{\phi}=0
\label{eq:3}
\een
where  
\ben
\tilde G^{\mu\nu}\equiv L_{X} g^{\mu\nu} + L_{XX} \nabla ^{\mu}\phi\nabla^{\nu}\phi
\label{eq:4}
\een
and $1+ {2X  L_{XX}\over L_{X}} > 0$.

Carrying out the conformal transformation
$G^{\mu\nu}\equiv {c_{s}\over L_{x}^{2}}\tilde G^{\mu\nu}$, with
$c_s^{2}(X,\phi)\equiv{(1+2X{L_{XX}\over L_{X}})^{-1}}\equiv sound ~ speed $,
the inverse metric of $G^{\mu\nu}$ is obtained as  
\ben G_{\mu\nu}={L_{X}\over c_{s}}[g_{\mu\nu}-{c_{s}^{2}}{L_{XX}\over L_{X}}\nabla_{\mu}\phi\nabla_{\nu}\phi] 
\label{eq:5}
\een
Another conformal transformation $\bar G_{\mu\nu}\equiv {c_{s}\over L_{X}}G_{\mu\nu}$ gives
\ben \bar G_{\mu\nu}
={g_{\mu\nu}-{{L_{XX}}\over {L_{X}+2XL_{XX}}}\nabla_{\mu}\phi\nabla_{\nu}\phi}
\label{eq:6}
\een	
$L_{X}\neq 0$ for the sound speed 
$c_{s}^{2}$ to be positive definite and only 
then equations $(1)-(4)$ will be physically meaningful. 
This is as as follows.   $L_{X}=0$ means that $L$ independent of  
$X$ so that in equation (\ref{eq:1}),  $L(X,\phi)\equiv L(\phi)$.
Then  $L$ becomes pure potential and 
the very definition of $k-$essence fields becomes meaningless because 
such fields correspond to lagrangians where kinetic energy dominates 
over potential energy. Also the very concept of minimal coupling of $\phi$ to 
$g_{\mu\nu}$ becomes redundant and 
equation (\ref{eq:1}) meaningless and equations (\ref{eq:4}-\ref{eq:6}) ambiguous.

For the non-trivial configurations 
of $\phi$ , $\partial_{\mu}\phi\neq 0$ and  $\bar G_{\mu\nu}$ is not conformally 
equivalent to $g_{\mu\nu}$. So this $\phi$ field has properties different 
from canonical scalar fields defined with $g_{\mu\nu}$ and the local causal 
structure is also different from those defined with $g_{\mu\nu}$.
Again, if $L$ is not an explicit function of $\phi$
then the equation of motion $(3)$ is replaced by ;
\ben
-{1\over \sqrt {-g}}{\delta S_{k}\over \delta \phi}
= \bar G^{\mu\nu}\nabla_{\mu}\nabla_{\nu}\phi=0
\label{eq:7}
\een
We take the  Lagrangian as $L=L(X)=1-V\sqrt{1-2X}$. 
This is a particular case of the BI lagrangian\\ 
$L(X,\phi)= 1-V(\phi)\sqrt{1-2X}$\\
 for $V(\phi)=V=constant$
and  $V<<kinetic ~ energy ~ of~\phi$ i.e.$V<< (\dot\phi)^{2}$. This 
is typical for the $k-$essence field where the kinetic energy 
dominates over the potential energy.
Then $c_{s}^{2}(X,\phi)=1-2X$.
For scalar fields $\nabla_{\mu}\phi=\partial_{\mu}\phi$. Thus (\ref{eq:6}) becomes
\ben
\bar G_{\mu\nu}= g_{\mu\nu} - \partial _{\mu}\phi\partial_{\nu}\phi
\label{eq:8}
\een
Note the rationale of using two conformal transformations:
the first is used to identify the inverse metric $G_{\mu\nu}$, while 
the second realises the mapping onto the   
metric given in $(8)$ for the lagrangian $L(X)=1 -V\sqrt{1-2X}$.

\section{The Reissner-Nordstrom case and mapping on to the  Robinson-Trautman type metric }
First consider the gravitational metric $g_{\mu\nu}$  to be Reissner-Nordstrom and denote
 $\partial_{0}\phi\equiv\dot\phi$, $\partial_{r}\phi\equiv\phi '$. Assuming 
that the $k-$ essence field $\phi (r,t)$ is spherically symmetric one has 
\ben
\bar G_{00}= g_{00} - (\partial _{0}\phi)^{2}=1-2GM/r+Q^{2} /r^{2}- \dot\phi ^{2}\nonumber\\
\bar G_{11}= g_{11} - (\partial _{r}\phi)^{2}= -(1-2GM/r+Q^{2} /r^{2})^{-1} - (\phi ') ^{2} \nonumber\\
\bar G_{22}= g_{22}=-r^{2}\nonumber\\ 
\bar G_{33}= g_{33}=-r^{2}sin^{2}\theta\nonumber\\
\bar G_{01}=\bar G_{10}=-\dot\phi\phi ' \nonumber\\ 
\label{eq:9}
\een
For the R-N metric,
$g_{00}=(1-2GM/r+Q^{2} /r^{2});g_{11}=-(1-2GM/r+Q^{2} /r^{2})^{-1};
g_{22}=-r^{2}; g_{33}=-r^{2}sin^{2}\theta; g_{ij} (i\neq j)=0$.

Note that the RN metric is spherically symmetric. The emergent gravity metric (\ref{eq:9}) 
contains additional terms {\it but all these are independent of $\theta$}. So the emergent 
metric is also spherically symmetric. So we might as well consider  $\theta=0$.
Then the emergent gravity line element becomes
\ben
ds_{RN,\theta=0}^{2}=(1-2GM/r+Q^{2} /r^{2} - \dot\phi ^{2})dt^{2}\nonumber\\
-((1-2GM/r+Q^{2} /r^{2})^{-1} + (\phi ') ^{2})dr^{2}\nonumber\\
-2\dot\phi\phi 'dtdr\label{eq:10}
\een
Making a co-ordinate transformation from $(t,r)$ to 
$(\omega,r)$ along $\theta=0$ such that (\cite{wein}) :
\ben
d\omega=dt-({{\dot\phi\phi '}\over{1-2GM/r+Q^{2} /r^{2} - \dot\phi ^{2}}})dr\label{eq:11}
\een
Then (\ref{eq:10}) becomes
\ben
ds^{2}=(1-2GM/r+Q^{2} /r^{2} - \dot\phi ^{2})d\omega^{2}\nonumber\\
-[{{(\dot\phi\phi ')^{2}}\over{(1-2GM/r+Q^{2} /r^{2}-\dot\phi^{2})}}\nonumber\\+{1\over(1-2GM/r+Q^{2} /r^{2})}+{(\phi ') ^{2}}]dr^{2}\nonumber\\
\nonumber\\
\label{eq:12}
\een
(\ref{eq:12}) will be a blackhole metric if $\bar G_{00}= \bar G_{11}^{-1}$ , i.e.
\ben
\dot\phi^{2}=(\phi')^{2}(1-2GM/r+Q^{2} /r^{2})^{2}
\label{eq:13}
\een
Let us assume a solution to (\ref{eq:13}) of the form 
$\phi(r,t)=\phi_{1}(r)+\phi_{2}(t)$.
Then (\ref{eq:13}) reduces to 
\ben
\dot\phi_{2}^{2}=(\phi_{1}')^{2}(1-2GM/r+Q^{2} /r^{2})^{2}= K
\label{eq:14}
\een 
$K(\neq 0)$ is a constant ($K\neq 0$ means $k-$essence field will have {\it non-zero} kinetic energy).
The solution to (\ref{eq:14}) 
\ben
\phi(r,t)=\phi_{1}(r)+\phi_{2}(t)\nonumber\\
=\sqrt{K}[r+{{(2G^{2}M^{2}-Q^{2})tan^{-1}{(r-GM)\over{\sqrt{Q^{2}-G^{2}M^{2}}})}}\over{\sqrt{Q^{2}-G^{2}M^{2}}}}
\nonumber\\+GM ln~(Q^{2}-2GMr+r^{2})]+\sqrt{K}t
\label{eq:15}
\een
with  $\phi_{1}(r)=\sqrt{K}[r+{{(2G^{2}M^{2}-Q^{2})tan^{-1}{(r-GM)\over{\sqrt{Q^{2}-G^{2}M^{2}}})}}\over{\sqrt{Q^{2}-G^{2}M^{2}}}}
\nonumber\\+GM ln~(Q^{2}-2GMr+r^{2})] ;$ and $\phi_{2}(t)=\sqrt{K}t$,
and we have taken an arbitrary integration constant to be zero.
Therefore the line element (\ref{eq:12}) becomes
\ben
ds^{2}=(1-{2GM\over r}+{Q^{2} \over r^{2}}-K)d\omega^{2}\nonumber\\
-{1\over(1-{2GM\over r}+{Q^{2}\over r^{2}}-K)}dr^{2}\nonumber\\
\label{eq:16}
\een
i.e.
\ben
ds^{2}=(\beta-{2GM\over r}+{Q^{2} \over r^{2}})d\omega^{2}
-{1\over(\beta-{2GM\over r}+{Q^{2}\over r^{2}})}dr^{2}\nonumber\\
\label{eq:17}
\een
with $\beta=1-K$. Now going over to the Eddington-Finkelstein coordinates $(v,r)$ or 
$(u,r)$ along $\theta=0$
i.e., {\it introducing advanced and retarded null coordinates}
\ben
v=\omega+r^{*}~~;~~u=\omega-r^{*}\nonumber\\
r^{*}= {1\over \beta}[r+{r_{+}^{2}\over{r_{+}-r_{-}}}ln~|r-r_{+}|-{r_{-}^{2}\over{r_{+}-r_{-}}}ln~|r-r_{-}|]
\nonumber\\
\label{eq:18}
\een
with $r_{+}={GM\over{\beta}}+{1\over{\beta}}\sqrt{(GM)^{2}-\beta{Q^{2}}}$ \nonumber\\
 and $r_{-}={GM\over{\beta}}-{1\over{\beta}}\sqrt{(GM)^{2}-\beta{Q^{2}}}$.

Then the line element (\ref{eq:17}) becomes
\ben
ds^{2}=(\beta-{2GM\over r}+{Q^{2} \over r^{2}})dv^{2}
-2dv dr\nonumber\\
\label{eq:19}
\een
or
\ben
ds^{2}=(\beta-{2GM\over r}+{Q^{2} \over r^{2}})du^{2}
+2du dr\nonumber\\
\label{eq:20}
\een
which is analogous to the {\it Robinson-Trautman} (RT) metric \cite{hans} along 
$\theta=0$ where $\beta$ can take the values $+1,0,-1$. In our case $\beta \neq +1$ because then  
$K=\dot\phi_{2}^{2}=0$ and dark energy is absent. $\beta\neq -1$, i.e. 
$K\neq 2$ as the total energy density 
cannot exceed unity ($\Omega_{matter} +\Omega_{radiation} +\Omega_{dark energy}= 1$).
 
{\it Therefore, the only allowed value of $\beta= 0$ i.e.$K=1$ and this is a perfectly valid solution 
because the RT metric allows $\beta=0$. Physically this means that $r_{+}$ is pushed to infinity while 
$r_{-}$ is pushed to zero.  This implies the radial coordinate r is a time-like 
coordinate on the whole space-time manifold and the outer horizon a sort of cosmological horizon.
Thus , as argued in reference \cite{schutz},
the case K=1 of (16) does not seem have a 
Newtonian limit, which makes it unsuitable for describing astrophysical objects.
However, although this may not 
be suitable as an astrophysical object but still is a consistent solution of Einstein's equation. In this 
context, it should be noted that even the Schwarzschild blackhole solution is strictly not astrophysically 
ever possible  because we cannot have static blackholes.  But still the Schwarzschild solution has been a 
milestone in understanding various nuances of general relativity. Similar situation prevails also for the 
Reissner-Nordstrom blackhole as charged blackholes are highly unlikely in nature for obvious reasons.}
So any confusion regarding $K$ taking the value $+1$ should 
not arise. We shall show below that $K=1$ gives zero Hawking temperature.

Also note that the solution $\phi(r,t)$ (\ref{eq:15}) 
{\it obtained from the blackhole conditions $\bar G_{00}= \bar G_{11}^{-1}$ } 
also satisfies the emergent gravity  equation of motion (\ref{eq:7}) at $r\rightarrow\infty$
along the symmetry axis, $\theta=0$:\\
$\bar G^{00}\partial_{0}^{2}\phi_{\mathrm 2} 
+ [\bar G^{11}(\partial_{1}^{2}\phi_{\mathrm 1} 
-\Gamma_{11}^{1}\partial_{1}\phi_{\mathrm 1})]
+\bar G^{01}\nabla_{0}\nabla_{1}\phi\nonumber\\ 
+\bar G^{10}\nabla_{1}\nabla_{0}\phi= 0$.
The first term vanish since $\phi_{2}(t)$ linear in $t$ and second term within third bracket vanish
at $r\rightarrow\infty$ because 
$2nd term =-\sqrt{K}Q^{2}{(\beta r^{2}-2GMr+Q^{2})\over{r(r^{2}-2GMr+Q^{2})^{2}}} $ and
 the last two terms vanish because $\bar G^{01}=\bar G^{10}=0$.

So the scalar field that one needs to produce an emergent RN black hole 
satisfies the equation of motion of emergent gravity (7) only for infinite coordinate radius along the 
polar axis. One may question what the geometry discussed has to do with emergent 
gravity in the first place. The answer is that as the emergent geometry has a scalar field intricately 
linked with it {\it a priori}, having a solution at $r\rightarrow\infty$ is non-trivial from various 
aspects. Let us discuss these.

First note that the solution for the scalar field $\phi$ ,(\ref{eq:15}), does not vanish for 
$r\rightarrow\infty$ as is usually expected of well behaved fields.
Here $\phi_{1}(r\rightarrow\infty)= {\sqrt K}[ r + 2GM ln r  
+  \frac {2G^{2}M^{2}-Q^{2}} {\sqrt{Q^{2}-G^{2}M^{2}}}\frac{\pi}{2}]$.    

Moreover, if $Q-GM=\alpha$ where $\alpha\rightarrow 0$, so that $ln r$ is negligible compared to the 
other terms 
then $\phi_{1}(r\rightarrow\infty)\sim {\sqrt K}[ r + \frac{\sqrt Q}{\sqrt 2\alpha}\frac{\pi}{2}]$. 
All these are solutions of the theory and so deserve mention.

\section{The Hawking Temperature for Robinson-Trautman type metric}

We use the tunnelling method to calculate the Hawking temperature for  (\ref{eq:19}) \cite{mitra}:
A massless particle in a black hole background is described by the Klein-Gordon equation
\ben
\hbar^2(-\bar G)^{-1/2}\partial_\mu( \bar G ^{\mu\nu}(-\bar G)^{1/2}\partial_\nu\Psi)=0.
\label{eq:21}
\een
One expands
\ben
\Psi=exp({i\over{\hbar}}S+...)
\label{eq:22}
\een
to obtain the leading order in $\hbar$ the Hamilton-Jacobi equation is
\ben
\bar G^{\mu\nu}\partial_\mu S \partial_\nu S=0
\label{eq:23}
\een
Assume S is independent of $\theta$ and $\phi$. Then 
\ben
2{\partial S\over{\partial v}}{\partial S\over{\partial r}}+
(\beta-{2GM\over{r}}+{Q^{2}\over{r^{2}}})({\partial S\over{\partial r}})^{2}=0
\label{eq:24}
\een
The symmetries of the metric permit the action to be written as
\ben
S=-Ev+W(r)+J(x^{i})
\label{eq:25}
\een
Then 
\ben
\partial_{v}S=-E~;~\partial_{r}S=W^{'}~;~\partial_{i}S=J_{i}
\label{eq:26}
\een
$J_{i}$ are constants chosen to be zero.
Combining equations (\ref{eq:24}) and (\ref{eq:26}):
\ben
-2EW^{'}(r)+(\beta-{2GM\over{r}}+{Q^{2}\over{r^{2}}})(W^{'}(r))^{2}=0
\label{eq:27}
\een
Thus
\ben
W(r)=2\pi i {E\over{\beta}}{r_{+}^{2}\over{r_{+}-r_{-}}}
+2\pi i {E\over{\beta}}{r_{-}^{2}\over{r_{-}-r_{+}}}\nonumber\\
=W(r_{+})+W(r_{-})
\label{eq:28}
\een
The two values of $W(r)$ correspond to the processes that the particle tunnels through the outer and
inner horizons respectively.

Therefore 
\ben
S=-Ev+2\pi i {E\over{\beta}}{r_{+}^{2}\over{r_{+}-r_{-}}}
+2\pi i {E\over{\beta}}{r_{-}^{2}\over{r_{-}-r_{+}}}+J(x^{i})\nonumber\\
\label{eq:29}
\een
The tunneling rates of the outer and inner horizons are
\ben
\Gamma^{RT}_{+emergent} \sim e^{-2Im S{+}} \sim e^{-2Im W(r_{+})}\nonumber\\
=e^{4\pi {E\over{\beta}}{r_{+}^{2}\over{r_{+}-r_{-}}}}=e^{-{E\over{K_{B}T_{+}}}}
\label{eq:30}
\een
\ben
\Gamma^{RT}_{-emergent} \sim e^{-2Im S{-}} \sim e^{-2Im W(r_{-})}\nonumber\\
=e^{4\pi {E\over{\beta}}{r_{-}^{2}\over{r_{-}-r_{+}}}}=e^{-{E\over{K_{B}T_{-}}}}
\label{eq:31}
\een
From these two equations the corresponding Hawking temperatures of the two horizons 
are respectively 
\ben
T_{+emergent}^{RT}={\hbar c^{3}(1-K)^{2}\over{2\pi k_{B}}} 
{\sqrt{G^{2}M^{2}-Q^{2}(1-K)}
\over{[GM+\sqrt{G^{2}M^{2}-Q^{2}(1-K)}]^{2}}}\nonumber\\
\label{eq:32}
\een
and
\ben
T_{-emergent}^{RT}=-{\hbar c^{3}(1-K)^{2}\over{2\pi k_{B}}}
{\sqrt{G^{2}M^{2}-Q^{2}(1-K)}
\over{[GM-\sqrt{G^{2}M^{2}-Q^{2}(1-K)}]^{2}}}\nonumber\\
\label{eq:33}
\een
Hence, as stated before, the Hawking temperature for this case will vanish as the dark energy density 
has to be $K=\dot\phi_{2}^{2}=1$. So this RT blackhole in presence of dark energy cannot radiate as the 
dark energy density is constrained to be unity. 

\section{Emergent gravity and  Kerr  metric}
Now take the gravitational metric $g_{\mu\nu}$ to be Kerr. 
The line element is
\ben
ds_{Kerr}^{2}=(1-{2GMr\over{\rho^{2}}})dt^{2}+{4GMr\alpha sin^{2}\theta \over{\rho^{2}}}d\phi dt
-{\rho^{2}\over{\Delta}}dr^{2}\nonumber\\
-\rho^{2}d\theta^{2}-(r^{2}+\alpha^{2}+{2GMr\alpha^{2}sin^{2}\theta \over{\rho^{2}}})sin^{2}\theta d\phi^{2}\nonumber\\
\label{eq:34}
\een
where, $\alpha={J\over{GM}} ~~;~~ \rho^{2}=r^{2}+\alpha^{2}cos^{2}\theta~~ and~~ 
\Delta=r^{2}-2GMr+\alpha^{2}$.

In this context an important point should be stressed. Note that the above metric 
(\ref{eq:34}) can be recast (for zero total charge) into the form given in reference \cite{mann}
where the identifications are provided below. 
\ben
ds^{2}=f(r,\theta)dt^{2}-\frac{dr^{2}}{g(r,\theta)}+2H(r,\theta) dt d\phi \nonumber\\
-K(r,\theta)d\phi^{2}-\Sigma(r,\theta)d\theta^{2}
\label{eq:35}
\een
where,
$f(r,\theta)=\frac{\Delta(r)-\alpha^{2}sin^{2}\theta}{\Sigma(r,\theta)}$;\\
$g(r,\theta)=\frac{\Delta(r)}{\Sigma(r,\theta)}$;\\
$H(r,\theta)=\frac{\alpha sin^{2}\theta(r^{2}+\alpha^{2}-\Delta(r))}{\Sigma(r,\theta)}$;\\
$K(r,\theta)=\frac{(r^{2}+\alpha^{2})^{2}-\Delta(r)\alpha^{2} sin^{2}\theta}{\Sigma(r,\theta)}sin^{2}\theta$;\\
$\Sigma(r,\theta)=r^{2}+\alpha^{2}cos^{2}\theta$;\\
$\Delta(r)=r^{2}+\alpha^{2}-2GMr$.

In \cite{mann} it has been elaborately shown how 
the Hawking temperature is independent of $\theta$ although the metric functions depend on $\theta$.
In our case the emergent metric $\bar G_{\mu\nu}$ contains additional terms {\it but these additional 
terms are still independent of $\theta$. Therefore, the modified Hawking temperature will still be independent 
of $\theta$. Therefore we might as well do our evaluation for some fixed $\theta$ , i.e. $\theta=0$.} 
We consider the Kerr metric along $\theta=0$.  
Then (\ref{eq:34}) becomes \cite{chandra}
\ben
ds_{Kerr;\theta=0}^{2}={\Delta \over{\rho^{2}}}dt^{2}-{\rho^{2}\over{\Delta}}dr^{2}
\label{eq:36}
\een
where, $\rho^{2}=r^{2}+\alpha^{2}~~and~~\Delta=r^{2}-2GMr+\alpha^{2}$.
It is to be noted that the same  metric (\ref{eq:36}) was rediscovered in \cite{wil} 
using a different route. 

As before, we take the $k-$essence field $\phi(r,t)$ to be  spherically symmetric
in keeping with the usual spherically symmetric Born-Infeld type of lagrangian 
for the $k-$essence scalar field. This does imply any  necessary conflict with the non-spherically 
symmetric background.
 
Then  one has from (\ref{eq:8})
\ben
\bar G_{00}=g_{00}-(\partial _{0}\phi)^{2}={\Delta \over{\rho^{2}}}- \dot\phi ^{2}\nonumber\\
\bar G_{11}= g_{11} - (\partial _{r}\phi)^{2}= -{\rho^{2}\over{\Delta}} - (\phi ') ^{2}\nonumber\\
\bar G_{01}=\bar G_{10}=-\dot\phi\phi '. 
\label{eq:37}
\een
The emergent gravity line element (\ref{eq:37}) along $\theta=0$ is now 
\ben
ds^{2}=({\Delta \over{\rho^{2}}}- \dot\phi ^{2})dt^{2}
-({\rho^{2}\over{\Delta}} + (\phi ') ^{2})dr^{2}-2\dot\phi\phi 'dtdr
\label{eq:38}
\een
Now make a coordinate transformation from $(t,r)$ to 
$(\omega,r)$ such that
\ben
d\omega=dt-({\dot\phi \phi ' \over{{\Delta \over{\rho^{2}}}- \dot\phi ^{2}}})dr
\label{eq:39}
\een
Then (\ref{eq:38}) becomes
\ben
ds^{2}=({\Delta \over{\rho^{2}}}-\dot \phi^{2})d\omega^{2}
-({(\dot \phi \phi ')^{2}\over{({\Delta \over{\rho^{2}}}-\dot \phi^{2})}}
+{\rho^{2}\over{\Delta}}+(\phi ')^{2})dr^{2}
\label{eq:40}
\een
This equation (\ref{eq:40}) will   a black hole metric if $\bar G_{00}= \bar G_{11}^{-1}$, i.e.
\ben
\dot\phi^{2}={\Delta \over{\rho^{2}}}+{\Delta^{2}\over{\rho^{4}}}(\phi ')^{2}-{1\over{\rho^{2}}}
\label{eq:41}
\een
We take a solution of (\ref{eq:41}) as
$\phi(r,t)=\phi_{1}(r)+\phi_{2}(t)$.

So (\ref{eq:41}) reduces to 
\ben
\dot\phi_{2}^{2}={\Delta \over{\rho^{2}}}+{\Delta^{2}\over{\rho^{4}}}(\phi_{1} ')^{2}
-{1\over{\rho^{2}}}=K
\label{eq:42}
\een
where, $K(\neq 0)$ is a constant ($K\neq 0$ means $k-$essence field will have {\it non-zero}
kinetic energy).
Now from (\ref{eq:42}) we get,
$\dot \phi_{2}=\sqrt{K}$
and $\phi_{1} '={(\sqrt{r^{2}+\alpha^{2}})(\sqrt{r^{2}(K-1)+\alpha^{2}(K-1)+2GMr+1})
\over({r^{2}-2GMr+\alpha^{2}})}$.

The solution of (\ref{eq:42}) is
\ben
\phi(r,t)=\phi_{1}(r)+\phi_{2}(t)\nonumber\\
=(\sqrt{r^{2}(K-1)+\alpha^{2}(K-1)+2GMr+1})\nonumber\\
({\sqrt{2GMr}~~tan^{-1}({r\sqrt{2GMr} \over{\sqrt{\alpha^{2}-2GMr}\sqrt{\alpha^{2}+r^{2}}}})
\over{\sqrt{\alpha^{2}-2GMr}}}\nonumber\\
+ln(2(r+\sqrt{\alpha^{2}+r^{2}}))
+\sqrt{K}t  
\label{eq:43}
\een
where
$\phi_{1}(r)\nonumber\\
=(\sqrt{r^{2}(K-1)+\alpha^{2}(K-1)+2GMr+1})\nonumber\\
({\sqrt{2GMr}~~tan^{-1}({r\sqrt{2GMr} \over{\sqrt{\alpha^{2}-2GMr}\sqrt{\alpha^{2}+r^{2}}}})
\over{\sqrt{\alpha^{2}-2GMr}}}+ln(2(r+\sqrt{\alpha^{2}+r^{2}})))$\\ 
and $\phi_{2}(t)=\sqrt{K}t$ and choosing an arbitrary integration constant to be zero.
Therefore the line elements (\ref{eq:40}) becomes
\ben
ds^{2}=({\Delta \over{\rho^{2}}}-K)d\omega^{2}-{1\over{({\Delta \over{\rho^{2}}}-K)}}dr^{2}\nonumber\\
=\frac{\beta\Delta'}{\rho^2}d\omega^2 - \frac {\rho^2}{\beta\Delta'} dr^2
\label{eq:44}
\een
where $\beta=1-K$, $M'=\frac {M}{1-K}$, $\Delta'= (r^{2}-2GM'r+\alpha^{2})$ and $\rho^2=r^2+\alpha^2$.
{\it Here note that $K\neq 1$ since $\beta$ cannot be zero, as then the metric 
becomes singular. $K$ cannot be greater than $1$ because then the signature of the metric 
(\ref{eq:44}) will be wrong. $K\neq 0$ because that would imply dark energy is absent. 
Therefore, the only allowed values are  $0 < K < 1$. So there is no question of $K$ approaching $1$ 
and confusions regarding this limit should not arise}. 
It can be shown that for $r\rightarrow\infty$ this metric is an approximate solution of 
Einstein's equations as the relevant terms fall of as $\frac {1}{r^{3}}$.

We now show that there is a further restriction on the dark energy density $K=\dot\phi_{2}^2$
if we want the fields $\phi (r,t)$ given by (\ref{eq:43}) to satisfy the equation of motion $(7)$ 
along the symmetry axis $\theta=0$ at $r \rightarrow \infty$.
For the axi-symmetric case, the equation of motion $(7)$ takes the form 
$\bar G^{00}\partial_{0}^{2}\phi_{\mathrm 2} 
+ \bar G^{11}\partial_{1}^{2}\phi_{\mathrm 1} 
-\bar G^{11}\Gamma_{11}^{1}\partial_{1}\phi_{\mathrm 1}
+\bar G^{01}\nabla_{0}\nabla_{1}\phi
+\bar G^{10}\nabla_{1}\nabla_{0}\phi= 0$. 
The first term vanishes exactly because $\phi_{2}(t)$ is linear in $t$, 
and the last two terms vanish because $\bar G^{01}=\bar G^{10}=0$. 

Using the expression for $\Gamma_{11}^{1}= \frac{GM (\alpha^2-r^2)}{(r^2+\alpha^2)(r^2-2GMr +\alpha^2)}$ 
the third term for $r\rightarrow\infty$ goes as $\frac{1}{r^2}$ and hence may be ignored. The remaining 
second term for $r\rightarrow\infty$ goes as $\frac{|1-K|^{{3\over 2}}}{r}$. 
As per the Planck collaboration results  \cite{planck}, the value of dark energy density $K$ is about $0.696$.
Therefore, the term is negligible as the denominator goes to infinity. Therefore 
in this limit this term also may be ignored and hence the equations of motion satisfied. 
{\it Therefore,   $K\neq 0,1$ and $0<K<1$. However $K$ should be  very close to unity for equations of motion 
to be satisfied at large $r$}.  

\section{The Hawking temperature for Kerr type metric}
Now we go over to the Eddington-Finkelstein coordinates $(v,r)$ 
or $(u,r)$ along the symmetry axis $\theta=0$.
$v=\omega+r^{*}$ and $u=\omega-r^{*}$ ,$\beta= 1-K $ and
\ben
r^{*}= \beta ^{-1}[r+({r_{+}^{2}+\alpha^{2}\over{r_{+}-r_{-}}})ln~|r-r_{+}|
-({r_{-}^{2}+\alpha^{2}\over{r_{+}-r_{-}}})ln~|r-r_{-}|]\nonumber\\
\label{eq:45}
\een  
with $r_{+}=GM'+\sqrt{(GM')^{2}-\alpha^{2}}={GM\over{1-K}}+\sqrt{({GM\over{1-K}})^{2}-\alpha^{2}}$\nonumber\\
and  $r_{-}=GM'-\sqrt{(GM')^{2}-\alpha^{2}}={GM\over{1-K}}-\sqrt{({GM\over{1-K}})^{2}-\alpha^{2}}$.
Therefore the line element (\ref{eq:44}) 
\ben
ds^{2}=({\beta\Delta' \over{r^{2}+\alpha^{2}}})dv^{2}-2dv dr
={\beta(r-r_{+})(r-r_{-}) \over{r^{2}+\alpha^{2}}}dv^{2}-2dv dr.\nonumber\\
\label{eq:46}
\een

Now proceeding exactly as before we calculate the Hawking temperatures for the 
two horizons. These are  
\ben
T_{+emergent}^{K}={\hbar c^{3}(1-K)^{2}\over{4\pi k_{B}}}
{\sqrt{(GM)^{2}-\alpha^{2}(1-K)^{2}}
\over{(GM)^{2}+GM\sqrt{(GM)^{2}-\alpha^{2}(1-K)^{2}}}}\nonumber\\
\label{eq:47}
\een
and
\ben
T_{-emergent}^{K}=-{\hbar c^{3}(1-K)^{2}\over{4\pi k_{B}}}\nonumber\\
({\sqrt{(GM)^{2}-\alpha^{2}(1-K)^{2}}
\over{(GM)^{2}-GM\sqrt{(GM)^{2}-\alpha^{2}(1-K)^{2}}}})\nonumber\\
\label{eq:48}
\een
where, $k_{B}$ is the Boltzmann constant.

\section{Conclusion}

In this work we have determined the Hawking temperatures for emergent gravity metrics having 
Reissner-Nordstrom and Kerr backgrounds.
We have shown that presence of dark energy modifies the Hawking temperatures. We first do the exercise 
for  the spherically symmetric Reissner-Nordstrom background metric along $\theta=0$. For $\theta=0$ 
the $k$-essence scalar field satisfies the emergent gravity equations of motion.In this case, 
the emergent metric can be mapped into a Robinson-Trautman blackhole. 
When the dark energy density is unity this blackhole has zero Hawking temperature i.e. it does not radiate. 
We next work with a Kerr background along $\theta=0$ again so that the emergent gravity equations of motion 
are again satisfied by the dark energy field.
The emergent blackhole metric satisfies Einstein's equations for large $r$ and always radiates.
Our analysis is done in the 
context of emergent gravity metrics having $k-$essence scalar fields $\phi$ with a Born-Infeld type 
lagrangian.In both cases the 
scalar field $\phi(r,t)=\phi_{1}(r)+\phi_{2}(t)$ also satisfies the emergent gravity equations of motion at  
$r\rightarrow\infty$ for $\theta=0$. 

Note that in reference \cite{dg} 
the resulting emergent metric had been identified with a topological defect \cite{bv}. 
In this work there are no defects in the emergent metrics. There are two event horizons and hence  two 
Hawking temperatures in both cases. But only one of these temperatures ,
{\it viz.},that corresponding to the outer horizon is observationally relevant.


\begin{thebibliography}{99}

\bibitem{dg}
D.Gangopadhyay and Goutam Manna, Euro.Phys.Lett. {\bf 100} 49001 (2012).

\bibitem{haw}
S. Hawking, Phys. Rev. Letters {\bf 26}, 1344 (1971);~~
L. Smarr, Phys. Rev. Lett. 30, {\bf 71} (1973);~~
J. Bardeen, B. Carter and S. Hawking, Comm. Math. Phys. {\bf 31}, 161 (1973);~~
S. Hawking, Nature (London) 248, {\bf 30} (1974).
S. Hawking, Commun. Math. Phys. {\bf 43}, 199 (1975).
J. Bekenstein, Phys. Rev. {\bf D7}, 2333 (1973); Phys. Rev. {\bf D9}, 3292 (1974);~~
G. Gibbons and S. Hawking, Phys. Rev. {\bf D15}, 2752 (1977);~~
S.W. Hawking, G.T. Horowitz and S.F. Ross, Phys. Rev. {\bf D51}, 4302 (1995);~~
Maulik K. Parikh and Frank Wilczek, Phys.Rev.Lett.{\bf 85}, 5042 (2000);~~
K.Murata and J.Soda, Phys.Rev.{\bf D74}, 044018 (2006);~~
R.Kerner and R.B.Mann, Class.Quant.Grav. {\bf 25}, 095014 (2008);~~
Zheng Ze Ma, Phys.Lett. {\bf B666} 376 (2008).

\bibitem{scherrer}
V.Gorini,A.Kamenschik and U.Moschella, Phys.Rev. {\bf D67} 063509 (2003);~~
V.Gorini,A.Kamenschik and U.Moschella and V.Pasquier ,arXiv:gr-qc/0403062 (2004);~~; 
L.Rizzi,S.Cacciatori,V.Gorini,A.Kamenschik and O.F.Piatella, Phys.Rev {\bf D82} 027301 (2010);~~
A.Y.Kamenschik,A.Tronconi and G.Venturi, Phys.Lett. {\bf B702} 191 (2011);~~
R.J. Scherrer, Phys.Rev.Lett.{\bf 93} 011301 (2004);~~
L.P.Chimento, Phys.Rev.{\bf D69} 123517 (2004);~~
D.Gangopadhyay and S. Mukherjee, Phys. Lett.{\bf B665} 121 (2008);~~
D.Gangopadhyay, Gravitation and Cosmology {\bf 16} 231 (2010);~~
D.Gangopadhyay and G. Manna, EPL, {\bf 100} 49001 (2012).
M.Born and L.Infeld,Proc.Roy.Soc.Lond {\bf A144}(1934) 425.
C.Armendariz-Picon, T.Damour and V.Mukhanov, Phys.Lett.{\bf B458} 209 (1999);~~
C.Armendariz-Picon, V.Mukhanov and P.J.Steinhardt, Phys.Rev.{\bf D63} 103510 (2001);~~
T.Chiba, T.Okabe and M.Yamaguchi, Phys.Rev.{\bf D62} 023511 (2000);~~
C.Armendariz-Picon and E.A.Lim, JCAP {\bf 0508} 007 (2005);~~
N.Arkani-Hamed, H.C.Cheng,M.A.Luty and S.Mukohyama, JHEP {\bf 05} 074 (2004);~~
N.Arkani-Hamed, P.Creminelli,S.Mukohyama and M.Zaldarriaga, JCAP {\bf 0404} 001 (2004);~~
R.R.Caldwell, Phys.Lett.{\bf B545} 23 (2002);~~
J.Callan, G.Curtis and J.M.Maldacena, Nucl.Phys.{\bf B513} 198 (1998);~~
A.D.Rendall, Class.Quant.Grav.{\bf 23} 1557 (2006).
G.W.Gibbons,  Nucl.Phys.{\bf B514} 603 (1998);~~
G.W.Gibbons, Rev.Mex.Fis.{\bf 49S1} 19 (2003);~~

\bibitem{babi}
M.Visser,C.Barcelo and S.Liberati, Gen.Rel.Grav. {\bf 34} 1719 (2002);~~
E.Babichev, V.Mukhanov and A.Vikman, JHEP {\bf 09}, 061 (2006);~~
E.Babichev,M.Mukhanov and A.Vikman, JHEP {\bf 0802} 101 (2008);~~
E.Babichev,M.Mukhanov and A.Vikman, WSPC-Proceedings, February 1, 2008.

\bibitem{wein}
S. Weinberg, {\it Gravitation and Cosmology}, Wiley Student Edition, John Wiley and Sons (Asia) Pte. Ltd., 2004

\bibitem{hans}
Hans Stephani {\it et~al.\/} {\it Exat Solutions of Einstein's Field Equations}, Second Edition,
Cambridge university Press, (2003);~~
J.B.Griffiths and J.Podolsky, {\it Exat Space-Times in Einstein's General Relativity}, 
Cambridge University Press, (2009).

\bibitem{schutz}
Bernard F.Schutz, {\it A first course in General Relativity, Ch.8, section 4}, 
Cambridge University Press, (1985). 

\bibitem{mitra}
P.Mitra, Phys.Lett. {\bf B648}, 240 (2007);~~
Bhramar Chatterjee, A.Ghosh and P.Mitra, Phys.Lett. {\bf B661}, 307 (2008);~~
Bhramar Chatterjee and P.Mitra, Phys.Lett. {\bf B675}, 640 (2008);~~
P.Mitra, {\it Black Hole Entropy}, [arXiv: 0902.2055];~~
E.T. Akhmedov, T.Pilling, A.de Gill and D. Singleton, Phys.Lett {\bf 666} 269 (2008);~~
E.T. Akhmedov, T.Pilling and D. Singleton,Int.Jour.Mod.Phys. {\bf D17}, 2453 (2008);~~
Z. Zhai and W. Liu, {\bf 325}, 63-67, Astrophys Space Sci (2010).

\bibitem{mann}
R.Kerner and R.B.Mann, Phys.Rev. {\bf D73}, 104010 (2006).

\bibitem{chandra}
S. Chandrasekhar, {\it The Mathematical Theory of Black Holes}, Oxford University Press, (1992)

\bibitem{wil}
S.Iso, H.Umetsu, F.Wilczek, Phys.Rev.Lett. {\bf 96} 151302 (2006); S.Iso, H.Umetsu, F.Wilczek, Phys.Rev.{\bf D74} 044017 (2006); 
T. Zhu, Gen.Rel.Grav. {\bf 44} 1525 (2012).

\bibitem{planck}
Planck 2013 results. I. Overview of products and scientific results, Planck collaboration, arXiv.1303.5062;\\
Planck 2013 results. XVI. Cosmological parameters,Planck collaboration, arXiv.1303.5076.

\bibitem{bv}
M.Barriola  and A.Vilenkin, Phys.Rev.Lett. {\bf 63} 341 (1989);~~
H.Yu, Nucl.Phys.{\bf B430} 427 (1994);~~
D.Gangopadhyay, R.Bhattacharyya and L.P. Singh, Grav.and Cosmo.{\bf 13} 1 (2007).
D.Bazeia,M.A. Gonzalez Leon, L.Losano and J.Mateos Guilarte, Euro.Phys.Lett.,
{\bf 93} 41001 (2011);~~P.Avelino,D.Bazeia,R.Menezes and J.G.G.S.Ramos,
Euro.Phys.Jour.C {\bf 71} 1683 (2011).

\end{thebibliography}
\end{document}